\documentclass[a4paper, 12pt, oneside]{article}

\usepackage{aapreprint}

\usepackage{graphicx,wrapfig,float,slashed,subcaption,bbold,bm}
\usepackage{amsmath,amssymb,epsfig,graphicx,xcolor}
\usepackage{epstopdf}
\usepackage{booktabs}
\usepackage{ragged2e}
\usepackage{mciteplus}
\usepackage{mathtools}

\usepackage{cancel}
\usepackage{ulem}

\usepackage{makecell}

 \usepackage{comment}

\newcommand{\nc}{\newcommand}
\nc{\non}{\nonumber}
\nc{\hc}{\hbox {H.c.}}
\nc{\noi}{\noindent}
\nc{\barx}{\bar{x}}
\nc{\pbarn}{\;\hbox {pb}}
\nc{\fbarn}{\;\hbox {fb}}

\nc{\hsp}{\hspace{0.5cm}}
\nc{\lsp}{\hspace{1cm}}
\nc{\Lsp}{\hspace{2cm}}
\nc{\LLsp}{\lsp\lsp}
\nc{\lra}{\longrightarrow}
\nc{\p}{\prime}
\nc{\sgn}{\text{sgn}}
\nc{\ph}{\varphi}
\nc{\op}{{\cal O}}

\nc{\beq}{\begin{equation}}  \nc{\eeq}{\end{equation}}
\nc{\bea}{\begin{eqnarray}}  \nc{\eea}{\end{eqnarray}}
\nc{\baa}{\begin{array}}     \nc{\eaa}{\end{array}}
\nc{\bit}{\begin{itemize}}   \nc{\eit}{\end{itemize}}
\nc{\ben}{\begin{enumerate}} \nc{\een}{\end{enumerate}}
\nc{\bce}{\begin{center}}    \nc{\ece}{\end{center}}
\nc{\bpm}{\begin{pmatrix}}   \nc{\epm}{\end{pmatrix}}
\nc{\bvt}{\begin{verbatim}}  \nc{\evt}{\end{verbatim}}

\def\lsim{\mathrel{\raise.3ex\hbox{$<$\kern-.75em\lower1ex\hbox{$\sim$}}}}
\def\gsim{\mathrel{\raise.3ex\hbox{$>$\kern-.75em\lower1ex\hbox{$\sim$}}}}

\def\udots{\mathinner{\mkern1mu\raise1pt\vbox{\kern7pt\hbox{.}}\mkern2mu\raise4pt\hbox{.}\mkern2mu\raise7pt\hbox{.}\mkern1mu}}

\def\ev{\;\hbox{eV}}
\def\kev{\;\hbox{keV}}
\def\mev{\;\hbox{MeV}}
\def\gev{\;\hbox{GeV}}

\def\dd{\mathrm d}

\newcommand{\Fig}[1]{Fig.~\ref{#1}}
\newcommand{\Eq}[1]{Eq.~(\ref{#1})}
\newcommand{\Eqs}[2]{Eqs.~(\ref{#1}) and (\ref{#2})}

\newcommand{\cft}{\mathrm{CFT}}
\newcommand{\sm}{\mathrm{SM}}
\newcommand{\dm}{\mathrm{DM}}
\newcommand{\gap}{\mathrm{gap}}
\newcommand{\ir}{\mathrm{IR}}

\allowdisplaybreaks

\newcommand\fverb{\setbox\fverbbox=\hbox\bgroup\verb}
\newcommand\fverbdo{\egroup\medskip\noindent%
			\fbox{\unhbox\fverbbox}\ }
\newcommand\fverbit{\egroup\item[\fbox{\unhbox\fverbbox}]}
\newbox\fverbbox

\preprint{\begin{flushright}
\end{flushright}}

\title{Conformal Freeze-In from Neutrino Portal} 
\author[a]{Sungwoo Hong,}
\author[b]{Maxim Perelstein,}
\author[b,c]{and Taewook Youn}
\affiliation[a]{Department of Physics, Korea Advanced Institute of Science and Technology \\ Daejeon 34141, Republic of Korea}
\affiliation[b]{Laboratory for Elementary Particle Physics \\ Cornell University, Ithaca, NY 14853, USA}
\affiliation[c]{School of Physics, Korea Institute for Advanced Study \\  Seoul 02455, Republic of Korea}
\emailAdd{sungwooh@kaist.ac.kr}
\emailAdd{m.perelstein@cornell.edu}
\emailAdd{taewook.youn@cornell.edu}

\abstract{We study a scenario where a dark sector, described by a Conformal Field Theory (CFT), interacts with the Standard Model through the neutrino portal. In this setup, conformal invariance breaks below the electroweak scale, causing the theory to transition into a confined (hadronic) phase. One of the hadronic excitations in this phase can act as dark matter. In the “Conformal Freeze-In” cosmological framework, the dark sector is populated through interactions with the Standard Model at temperatures where it retains approximate conformal symmetry. The dark matter relic density depends on the CFT parameters, such as the dimension of the operator coupled to the Standard Model. We demonstrate that this model can reproduce the DM relic density and meet all observational constraints. The same neutrino portal interaction may also generate masses for the active neutrinos. The dark matter candidate could either be a pseudo-Goldstone boson (PGB) or a composite fermion with the quantum numbers of a sterile neutrino. In the latter case, the model is consistent with the current X-ray constraints, and may be detectable with future X-ray observations.}

\begin{document}

\maketitle
\flushbottom

\section{Introduction}
The existence of dark matter and massive neutrinos provides compelling evidence for physics beyond the Standard Model (BSM). Although the Standard Model (SM) successfully explains a wide range of phenomena, it does not account for the dark matter (DM) that makes up approximately 27\% of the universe. Additionally, the tiny masses of neutrinos, as indicated by oscillation experiments, suggest the presence of new physics. 
These two unresolved aspects of modern physics hint at a possible connection, making the study of dark matter and neutrino interactions a fertile ground for exploring BSM scenarios. Various theoretical models have been proposed in this context, including sneutrino dark matter in supersymmetric extensions of the SM~\cite{Ibanez:1983kw,Hagelin:1984wv,Hall:1997ah}, sterile neutrino dark matter~\cite{Dodelson:1993je,Shi:1998km,Abazajian:2001nj}, and scotogenic models that link neutrino masses to dark matter~\cite{Ma:2006km,Boehm:2006mi,Hambye:2009pw}.

One promising approach to understanding this connection lies in the framework of a ``dark sector", whose structure and complexity are comparable to those of the visible sector~\cite{Dienes:2022zbh,Essig:2013lka}. 
This hypothetical sector might communicate with the visible sector through a ``neutrino portal". This setup provides a minimal and theoretically appealing mechanism as the neutrino portal can produce dark matter and/or account for the small mass of neutrinos. For earlier studies on the neutrino portal, see Refs.~\cite{Escudero:2016ksa,Batell:2017cmf,Lamprea:2019qet}.

In this work, we consider a scenario in which the dark sector possesses an approximate conformal symmetry. Conformal field theories (CFTs) are common in the landscape of quantum field theories, emerging whenever the renormalization group evolution reaches a non-trivial attractive fixed point~\cite{DiFrancesco:1997nk,Ginsparg:1988ui, Rychkov:2016iqz}.
Conformal symmetry dictates that the energy density of the dark sector will scale like radiation in an expanding universe. This radiation-like scaling suggests that dark matter cannot be composed of CFT states. To incorporate DM, the conformal symmetry must be broken at a low-energy scale $M_\gap$. Below this scale, the theory transitions into a ``hadronic" phase, where ordinary massive particle excitations emerge in the spectrum. These particles can then serve as candidates for Cold Dark Matter (CDM). Furthermore, the small neutrino mass can arise through couplings to the composite states~\cite{Arkani-Hamed:1998wff,vonGersdorff:2008is,Grossman:2010iq, Agashe:2015izu, Agashe:2016ttz, Agashe:2017ann, Chacko:2020zze, Agashe:2024uvp}.

However, thermal equilibrium with the SM would require a mass gap in the range of 10 to 100 eV to match the observed relic density \cite{Hong:2022gzo}~\footnote{This conclusion can be circumvented if dark matter can efficiently annihilate into Standard Model particles during the hadronic phase. Ref.~\cite{Ahmed:2023vdb} discusses such a class of models with the neutrino portal.}; such a mass range is already excluded by constraints from large-scale structure. Instead, we explore a scenario in which the CFT sector remains out of thermal equilibrium with the visible sector because of a feeble coupling between them, and the very weak interaction with SM serves as the primary mechanism of populating the CFT sector in the early universe. This type of non-thermal production mechanism, similar to the ``freeze-in” process for ordinary particles, can be described as ``conformal freeze-in" (COFI)~\cite{Hong:2019nwd,Hong:2022gzo,Chiu:2022bni}.~\footnote{Recently, a mechanism for baryon asymmetry generation based on the conformal freeze-in was proposed in \cite{Agashe:2024uvp}.}

In this paper, we present an analysis of the COFI mechanism via the neutrino portal, or fermionic COFI. We find that this COFI scenario can account for both the origin of dark matter and the small neutrino masses, where the right-handed neutrino can be either composite or elementary. The potential dark matter candidates in this framework include a stable pseudoscalar or a metastable fermion. Notably, this metastable fermion has the quantum numbers of a sterile neutrino, and we find that such a sterile neutrino-like dark matter candidate can evade the stringent constraints imposed by X-ray experiments. This is in contrast with the sterile neutrino DM produced by the conventional Dodelson-Widrow mechanism~\cite{Dodelson:1993je}, which is currently ruled out. At the same time, the signatures of metastable COFI dark matter may be detected in future X-ray observations, providing a realistic observational signature of this scenario.   

The outline of the paper is as follows. In Section~\ref{sec:review}, we provide a brief overview of the COFI mechanism~\cite{Hong:2019nwd,Hong:2022gzo,Chiu:2022bni}. Sections~\ref{sec:fCOFI} and \ref{sec:relic} detail the fermionic COFI framework and its cosmological evolution and relic density, respectively. In Section~\ref{sec:nu_mass}, we discuss the generation of neutrino mass within this model. In Section~\ref{sec:pheno}, we conduct a comprehensive phenomenological analysis of dark matter. Finally, we conclude in Section~\ref{sec:con}.

\section{Brief Review of Conformal Freeze-In}
\label{sec:review}

\subsection{Conformal Dark Sector}
In the COFI scenario, the dark sector is described by a Conformal Field Theory (CFT) across a wide range of energy scales, between the ``gap scale" $M_\gap$ in the infrared (IR), and the ultraviolet (UV) cutoff\footnote{Above $\Lambda_{\rm UV}$, the dark sector could for example be described by a gauge theory flowing towards a strong-coupling analog of Banks-Zaks fixed point. See \cite{Hong:2019nwd,Hong:2022gzo} for more details.} $\Lambda_\text{UV} \gg M_\gap$. We assume that the dark CFT contains a relevant scalar operator $\mathcal O_\cft$ with scaling dimension $d < 4$.
In general, the CFT is strongly coupled and the scaling dimension $d$ is not necessarily an integer. This relevant scalar operator plays the role of a deformation to the CFT once added to the Lagrangian, and the conformal symmetry is generically broken in the IR. For example, if Lagrangian is given by
\beq
\mathcal L_\mathcal{O} \supset c \; \mathcal O_\cft, 
\label{eq:lagco}
\eeq
a Naive Dimensional Analysis (NDA) estimate suggests that the conformal symmetry is broken at the gap scale
\beq
M_\gap \sim c^{1/(4-d)}.
\eeq
Due to the uncertainties in the strongly-coupled theory, NDA estimates suffice for our purposes.

Below the gap scale $M_\gap$, the theory is no longer conformal, and the dark sector dynamics is assumed to be described in terms of hadronic composite states formed from the original degrees of freedom of the CFT.\footnote{While we consider the confinement phase as the IR phase of our theory, other possibilities exist, including the gapped continuum~\cite{Cabrer:2009we, Falkowski:2008fz, Stancato:2008mp, Falkowski:2008yr, Bellazzini:2015cgj, Csaki:2018kxb, Megias:2019vdb, Megias:2021mgj, Csaki:2021gfm, Csaki:2021xpy, Csaki:2022lnq, Ferrante:2023fpx}.} 
While predicting the spectrum of these particle-like excitations is highly challenging due to the nature of strong dynamics, we can make a realistic assumption that the lightest CFT composite state, denoted as $\chi$, is stable over cosmological timescales and serves as dark matter. This stability could be due to a conserved global symmetry carried by $\chi$ but not by any of the SM states.

In the existing COFI models \cite{Hong:2019nwd,Hong:2022gzo,Chiu:2022bni}, the dark matter particle is a pseudo-Goldstone boson (PGB) arising from an approximate global symmetry that is spontaneously broken at the scale $M_\gap$, similar to pions in ordinary QCD. In this scenario, it is natural to have $m_\dm \ll M_\gap$, necessary to meet self-interaction constraints \cite{Markevitch:2003at,Randall:2008ppe,Robertson:2016xjh}. Of course, any lightest and (pseudo-)stable composite state in the hadronic spectrum can be a dark matter candidate, including a composite fermion with the quantum numbers of a sterile neutrino. We will consider this possibility in this paper.  

The COFI mechanism considers a coupling between the SM and the dark sector of the form
\beq
\mathcal L_\mathcal{O} \supset \frac{\lambda_\cft}{\Lambda_\cft^{D-4}} \mathcal O_\sm \mathcal O_\cft,
\label{eq:osmocft}
\eeq
where $\mathcal O_\sm$ is a gauge-invariant operator consisting of the SM fields, $\lambda_\cft$ is a dimensionless coupling, and $\Lambda_\cft$ is (approximately) the energy scale where the above interaction is generated. Here $\mathcal O_\sm$ has a scaling dimension $d_\sm$, and we have
\beq
D = d_\sm + d.
\eeq
Note that the interaction \Eq{eq:osmocft} explicitly breaks the conformal symmetry, providing a source of CFT deformation. The simplest version of the COFI scenario assumes that this is the {\it only} (or, at least, the dominant) source of conformal symmetry breaking in the IR. 
If $\mathcal O_\sm$ acquires a non-zero vacuum expectation value (VEV), we are led to an effective Lagrangian of the form \Eq{eq:lagco}, with $c$ replaced by
\beq
c = \frac{\lambda_\cft}{\Lambda_\cft^{D-4}} \langle \mathcal{O}_\sm \rangle.
\label{eq:c_deform}
\eeq
This happens, for example, in the case of the Higgs portal operator, $\mathcal{O}_\sm=|H|^2$. If $\mathcal O_\sm$ does not acquire a vev, conformal symmetry breaking first arises at the one-loop level. One possibility is that loop diagrams generate
\beq
\mathcal L_\mathcal{O} \supset c' \mathcal O_\cft^2.
\label{eq:o2cft}
\eeq
This effect breaks the conformal symmetry in the IR when $\mathcal O^2_\cft$ is a relevant operator. In a large-$N$ approximation, this corresponds to $d < 2$. A second possibility is that loop diagrams induce an operator of the form $\sim \mathcal O_\sm^\prime \mathcal O_\cft$ with a different SM operator $O_\sm^\prime$ with its VEV non-zero. With such an induced operator, conformal symmetry breaking follows essentially the same steps as described above.

Previously, the COFI scenario has been studied with scalar interactions \cite{Hong:2019nwd,Hong:2022gzo} of the form
\beq
(H^\dagger H) \mathcal O_\cft, \quad (H Q^\dagger_L q_R) \mathcal O_\cft, \quad (G_{\mu\nu}G^{\mu\nu}) \mathcal O_\cft, \cdots
\eeq
and tensor interaction \cite{Chiu:2022bni} of the form
\beq
B_{\mu\nu} \mathcal O_\cft^{\mu\nu}.
\eeq
In this work, we consider a fermion interaction of the form
\beq
(HL)\mathcal O_\cft,
\eeq
which, as we discuss in detail below, can simultaneously generate neutrino mass and DM abundance via the COFI mechanism. Note that $HL$ is the only renormalizable gauge-invariant operator of spin 1/2 in the SM. 

\subsection{Cosmological Thermal History}

Similar to the usual freeze-in scenario~\cite{Hall:2009bx}, we assume that the inflaton only couples to the SM, so that the SM sector is reheated to temperature $T_R$ at the end of inflation, while the temperature of the dark sector remains low. The COFI scenario assumes
\beq
M_\gap < T_R < \Lambda_\cft.
\eeq
After reheating, collisions among the SM particles gradually populate the dark sector, and this occurs in the conformal phase if the energy of the colliding SM particles is larger than $M_\gap$. While zero temperature cross sections with CFT final states can be calculated by the “unparticle” technique of Georgi~\cite{Georgi:2007ek,Grinstein:2008qk}, thermal averaged rates, such as the energy transfer rates, and Boltzmann equations can be obtained by the method developed in \cite{Hong:2019nwd,Hong:2022gzo}.
Strong interactions among the CFT states effectively distribute the energy transferred into the conformal dark sector, creating a thermal CFT state. However, the interaction between SM and CFT sectors are sufficiently weak and two sectors do not reach a thermal equilibrium. This implies that the CFT temperature is always below that of the SM sector. After the dark sector goes through a confining phase transition, the energy stored in the CFT states is converted into the dark hadronic composite states, which then decay rapidly into stable dark matter. 

The evolution of the SM energy density $\rho_\sm$ is governed by the usual Boltzmann equations:
\beq
\frac{\dd \rho_\sm}{\dd t} + 3H(\rho_\sm + P_\sm) = -\Gamma_E(\sm\to\cft) + \Gamma_E(\cft\to\sm),
\label{eq:drhosm}
\eeq
where $H$ is the Hubble expansion rate, and $\Gamma_E (i \to f)$ are energy transfer rates. In the COFI scenario, we always have $\Gamma_E(\sm\to\cft) \gg \Gamma_E(\cft\to\sm)$, so the ``backreaction" rate $\Gamma_E(\cft\to\sm)$ will be ignored hereafter. The energy transfer rate from the SM sector is given by
\begin{equation}
\begin{aligned}
& \Gamma_E(\sm\to\cft)  = \sum_i n_i \langle \Gamma(i\to\cft)E\rangle + \sum_{i,j} n_i \langle \Gamma(i\to\cft+j)E\rangle \\
 & \;\;\; +\sum_{i,j} n_i n_j \langle \sigma(i+j\to\cft)v_\text{rel}E\rangle + \sum_{i,j,k} n_i n_j \langle \sigma(i+j\to\cft+k)v_\text{rel}E\rangle + \cdots\,,
\end{aligned}
\end{equation}
where the dots denote terms suppressed by powers of perturbative SM couplings and/or multi-particle phase space factors. 

Conformal symmetry of the dark sector ensures that the energy-momentum tensor of the CFT states is traceless, $P_\cft = \frac13 \rho_\cft$, and their energy density scales as radiation $\rho_\cft \propto a^{-4}$. Because the dark sector is populated during the radiation-dominated era, we have $P_\sm = \frac{1}{3}\rho_\sm$ and
\beq
\frac{\dd}{\dd t} (\rho_\cft + \rho_\sm)  + 4H(\rho_\cft + \rho_\sm) = 0.
\label{eq:rhosmcft}
\eeq
From \Eqs{eq:drhosm}{eq:rhosmcft}, we find that the CFT energy density evolves according to
\beq
\frac{\dd \rho_\cft}{\dd t} + 4H\rho_\cft = \Gamma_E(\sm\to\cft),
\label{eq:btzm}
\eeq
whose initial condition is given by $\rho_\cft = 0$ at $T_\sm = T_R$. 

By dimensional analysis, the energy transfer rate from the interaction \Eq{eq:osmocft} scales as
\beq
\Gamma_E(\sm) \sim \frac{\lambda^2_\cft}{\Lambda_\cft^{2(D-4)}} T^{2D-3}_\sm
\label{eq:gamD}
\eeq
when $T_\sm$ far exceeds all relevant SM energy scales and the mass gap of the dark sector. Solving \Eq{eq:btzm} further yields
\beq
\rho_\cft \propto T^4 \times \left( \frac{T^{2D-9}_R - T^{2D-9}}{2D-9} \right),
\eeq
which implies that $D =  d_\sm + d$ determines when the dark sector is mostly populated. Specifically, for $D < 4.5$, most of the dark sector energy density is generated at low temperatures (IR-dominant production), whereas for $D > 4.5$, the majority of the dark sector energy density is produced shortly after reheating (UV-dominant production). In the UV-dominant production, dark matter relic density does depend on the reheating temperature $T_R$.

The Boltzmann equation \Eq{eq:btzm} applies only when $T_\sm > M_\gap$ and $T_D > m_\dm$. When dark matter becomes non-relativistic, that is, $T_D < m_\dm$, $4H$ in the Hubble friction term should be replaced by $3H$ since the dark sector energy density evolves as non-relativistic matter rather than radiation. In most cases, the relevant SM particles drop out of the thermal bath at $T_\sm > M_\gap$, but light fields such as electrons or neutrinos can still generate dark matter for $M_\gap > T_\sm > m_\dm$. DM production in this regime is described by conventional freeze-in, and can be easily calculated. However, it was found in Ref.~\cite{Hong:2022gzo} that in the relevant parts of parameter space, dark matter production during the hadronic phase is typically insignificant compared to production in the CFT regime.

\section{Fermionic COFI Model}
\label{sec:fCOFI}

We will now study the conformal freeze-in production in a model where the coupling between the SM and the dark CFT is of the form
\beq
\mathcal L_{\mathcal O} = \frac{\lambda_\cft^{\alpha\beta}}{\Lambda_\cft^{d - 3/2}} (HL_\alpha) \mathcal O_\cft^\beta,
\label{eq:hln}
\eeq
where $\mathcal O_\cft$ is a spin-1/2 operator with the scaling dimension $d$. Note that unitarity bound in this case is $d \geq 3/2$. Here $\alpha, \beta$ are lepton flavor indices, which will be suppressed unless explicitly stated otherwise. The coupling constant $\lambda_\cft$ is dimensionless, and a flavor-diagonal and universal scheme is assumed for simplicity\footnote{This assumption can be naturally realized through extra-dimensional setups; for example, see Ref.~\cite{Desai:2020rwz}}. One of the goals of this work is to explain the origin of dark matter by the COFI mechanism generated by the $(HL) \mathcal O_\cft$ operator. 

\subsection{Conformal Symmetry Breaking}

\begin{figure}[t]
\centering
  \includegraphics[width=0.3\linewidth]{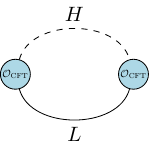}
\caption{Diagram that contributes to conformal symmetry breaking via generation of ${\cal O}^2_\cft$ from the $(HL) \mathcal O_\cft$ operator.}
\label{fig:mgap_diag}
\end{figure}

Conformal symmetry is broken in the SM, and the portal interaction in \Eq{eq:hln} will inevitably mediate this breaking to the dark sector.  
However, the SM $HL$ operator does not get a VEV. Being a spin-1/2 operator, it also cannot mix linearly with any operator that does acquire a VEV. As a result, the leading conformal symmetry-breaking effect in the dark sector due to the portal interaction arises at the one-loop level, see Fig.~\ref{fig:mgap_diag}. This can be represented by adding a term to the Lagrangian,
\beq
{\cal L} \supset C \mathcal O_\cft^2,
\label{eq:square}
\eeq
where $C$ is a constant that can be calculated from the diagram in Fig.~\ref{fig:mgap_diag}. We assume that the dark-sector CFT can be treated in the large-$N$ approximation, in which case the scaling dimension of ${\cal O}^2_\cft$ is approximately given by twice the scaling dimension of $\cal O_\cft$. There are two cases:

\begin{itemize}

\item $3/2\leq d < 2$: The deformation \Eq{eq:square} is relevant, and leads to the breaking of conformal symmetry in the infrared. The scale of this breaking is estimated as 
\beq
M^{(HL)}_\gap \sim C^{1/(4-d)}\,\sim\,\left( \frac{\lambda_\cft}{\Lambda_\cft^{d-3/2}} \frac{\Lambda_\sm^{1/2}}{4\pi} \right)^{\frac{1}{2-d}},
\label{eq:hlogap}
\eeq
where $\Lambda_\sm$ is the UV cut-off scale for the SM sector. Unfortunately, the calculation of DM relic density in the next section shows that if this effect is the leading source of conformal symmetry breaking in the dark sector, the resulting DM candidate is too light to be a viable DM candidate, due to warm dark matter constraints. Thus, while this effect is always there, an {\it additional}, stronger breaking of the conformal symmetry in the IR must be introduced.

\item $d\geq 2$: The deformation \Eq{eq:square} is marginal or irrelevant, and does not lead to the breaking of conformal symmetry in the infrared. Since such a breaking is a necessary condition to obtain a DM candidate at the end of the freeze-in, we again conclude that a new source of breaking of the conformal symmetry in the IR must be introduced.

\end{itemize}

Motivated by the above considerations, we assume that the CFT contains an additional relevant scalar operator $\tilde{\mathcal O}_\cft$ that generates the required mass gap: 
\beq
\mathcal L \supset \frac{\lambda_\cft}{\Lambda_\cft^{d - 3/2}} (HL) \mathcal O_\cft + \tilde{c} \, \tilde{\mathcal O}_\cft.
\label{eq:extra}
\eeq
In this setup, the mass gap is given by $M_\gap \sim \tilde{c}^{1/(4-\tilde{d})}$ and is an additional free parameter. This provides a phenomenologically viable fermionic COFI model, at the expense of losing some of the predictive power of the minimal bosonic COFI models discussed in Refs.~\cite{Hong:2019nwd,Hong:2022gzo} (see \cite{Chiu:2022bni} for the mass gap generation in the tensor COFI model where $M_\gap$ is induced from the dominant scalar operator in the operator product expansion $\cal O_{\mu\nu} \times \cal O^{\mu\nu}$). On the other hand, as we will see below, the fermionic COFI models also offer an important new advantage: the measured neutrino masses can be due to the same portal interaction, \Eq{eq:hln}, that generates the observed DM abundance.    

\subsection{Dark Matter Candidate}
\label{ssec:DMC}

Once the dark-sector conformal symmetry is broken at $M_\gap$, the CFT operator $\mathcal O_\cft$ is interpolated as an infinite tower of composite fermions operators $\psi_n$:
\beq
\mathcal O_\cft \sim \sum_n a_n M_\gap^{d-3/2} \psi_n, 
\label{eq:cftint}
\eeq
where the coefficients $a_n$ are model-dependent. If the lightest state $\psi_0$ in the $\psi_n$ tower is also the lightest in the hadron spectrum, $\psi_0$ may serve as a dark matter candidate. Since the quantum numbers of $\mathcal O_\cft$ are identical to those of the sterile neutrino, this model would provide a realization of familiar sterile neutrino DM, but with a novel production mechanism distinct from the usual Dodelson-Widrow (DW) mechanism~\cite{Dodelson:1993je}, where the sterile neutrino DM is produced via active-sterile neutrino oscillations. It should be noted that a sterile neutrino DM candidate cannot carry a symmetry charge not carried by the SM fields (this would be inconsistent with the assumed portal interaction,~\Eq{eq:hln}). Therefore, it cannot be absolutely stable, and indeed the decay $\psi_0 \to \nu +\gamma$ will provide a major constraint on this model.   

Depending on the details of the confining dynamics, other light composite states can also be dark matter candidate. For example, just like in bosonic COFI scenarios~\cite{Hong:2019nwd,Hong:2022gzo,Chiu:2022bni}, the DM particle might be a pseudo-Goldstone boson (PGB) of an approximate global symmetry spontaneously broken at $M_\gap$. This possibility is less constrained phenomenologically, since the PGB can carry an exact\footnote{As always, global symmetries are only exact up to Planck-suppressed operators. In general, coefficients of leading Planck-suppressed operators must be suppressed relative to their natural values to ensure DM stability, leading to the ``DM quality problem". In the COFI scenario, the quality problem is relatively mild due to the low mass scale of the DM, and only dimension-5 operators require significant suppression.} global symmetry. 

For concreteness, in the rest of this paper we will consider the dark matter particle to be either a fermion (composite sterile neutrino), or a PGB. We discuss the phenomenology in each of these two models in Section~\ref{sec:pheno}. In both cases, we will refer to the DM particle as $\chi$ in the rest of the paper.  

\section{Cosmological Evolution and Relic Density}
\label{sec:relic}

We assume that the reheating temperature $T_R$ is above the electroweak scale, and that only the SM gets reheated ({\it i.e.}~the inflaton does not couple directly to the dark sector). The energy transfer from the SM to the CFT is mediated by the portal interaction \Eq{eq:hln}. In this section we will compute the rate of this energy flow, and then use it to calculate the DM relic density in today's universe.

\subsection{Energy Transfer Rate}
\begin{figure}[h!]
	\centering
 	\includegraphics[width=.32\linewidth]{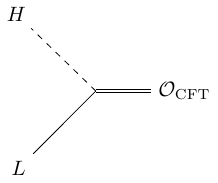}
	\includegraphics[width=.32\linewidth]{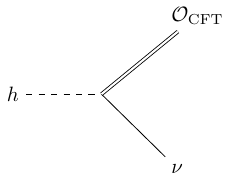}
	\includegraphics[width=.32\linewidth]{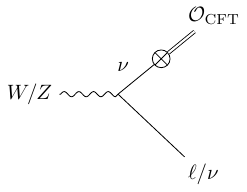}
	\caption{Feynman diagrams for dominant COFI processes. 
 }
	\label{fig:fyn}
\end{figure}

For SM temperatures above the weak scale, the $H+L \to \cft$ scattering process is the leading contribution to energy transfer\footnote{This process remains active below the weak scale, appearing in the form of $h + \nu \to \cft$ and $W + \ell \to \cft$, and it ends when relevant particles exit the thermal bath.}. Below the weak scale, the dominant production channels are decays of the Higgs or weak gauge bosons: $h \rightarrow \cft + \nu$ and $W/Z \rightarrow \cft + \ell/\nu$. The leading-order Feynman diagrams for these three processes are shown in Figure~\ref{fig:fyn}. Note that the gauge boson decay contains a mixing between $\nu$ and $\mathcal O_\cft$. 
This mixing is a new feature that was not present in the previously considered versions of the COFI scenario, and will lead to important differences between the fermionic and scalar/tensor cases. Even after the Higgs and $W/Z$ fall out of thermal bath, energy transfer to the CFT sector can continue via processes with multiple quarks and leptons, but these are subdominant due to suppression from extra phase space factors and gauge/Yukawa couplings. 

The energy transfer rate via the $H+L \to \cft$ process is given by
\begin{equation}
\begin{aligned}
&n_H n_L \langle \Gamma (H + L \rightarrow \cft) E\rangle \\
&\hskip 4em = \iiint \dd \Pi_H \dd \Pi_L \dd \Pi_\cft f_H f_L (2\pi)^4 \delta^4 (p_H + p_L - p_\cft) (E_H + E_L) |\mathcal M_{HL}|^2,
\end{aligned}
\end{equation}
while the energy transfer rate due to the Higgs and $W/Z$ decays can be computed as
\begin{equation}
\begin{aligned}
&n_h \langle \Gamma (h \rightarrow \cft + \nu) E\rangle \\
&\hskip 4em = \iiint \dd \Pi_h \dd \Pi_\nu \dd \Pi_\cft f_h (2\pi)^4 \delta^4 (p_h - p_\cft - p_\nu) (E_h - E_\nu) |\mathcal M_h|^2
\end{aligned}
\end{equation}
and
\begin{equation}
\begin{aligned}
&n_{W/Z} \langle \Gamma (W/Z \rightarrow \cft + \ell/\nu) E\rangle \\
&\hskip 4em \hspace{-1.5cm} = \iiint \dd \Pi_{W/Z} \dd \Pi_{\ell/\nu} \dd \Pi_\cft f_{W/Z} (2\pi)^4 \delta^4 (p_{W/Z} - p_\cft - p_{\ell/\nu}) (E_{W/Z} - E_{\ell/\nu}) |\mathcal M_{W/Z}|^2
\end{aligned}
\end{equation}
respectively. Here $f_i$'s are the thermal distribution functions, and $p_\cft$ is the 4-momentum carried by the conformal dark sector states. The CFT phase space is given by~\cite{Georgi:2007ek, Hong:2019nwd, Hong:2022gzo, Chacko:2020zze} 
\beq
\dd \Pi_\cft = \frac{\dd^4 p_\cft}{(2\pi)^4} A_{d-1/2} \theta(p_\cft^0) \theta(p_\cft^2 - \mu_\ir^2) (p_\cft^2 - \mu_\ir^2)^{d-5/2},
\eeq
where 
\beq
A_{d-1/2} = \frac{16 \pi^{5/2}}{(2\pi)^{2d-1}} \frac{\Gamma(d)}{\Gamma(d-3/2) \Gamma(2d-1)},
\eeq
and 
$\mu_\ir$ is the IR cutoff that reflects the breakdown of conformal invariance at the compositeness scale. For the rest of this paper, we will identify $\mu_\ir$ with the mass gap $M_\gap$. It may be worth mentioning that the IR cutoff should be thought of as a simple, crude way to model the complicated non-perturbative dynamics occurring at the gap scale. In situations where the integration over the CFT phase space is dominated by the IR part of the spectrum, our calculations should provide qualitatively reliable results, but are subject to order-one corrections from the region around $M_\gap$. In cases where the integration is dominated by regions away from the IR boundary, the calculations should be quantitatively reliable.   

The spin-averaged scattering amplitude for the $H+L\to\cft$ process is given by
\beq
\sum |\mathcal M_{HL}|^2 = \frac{\lambda^2_\cft}{\Lambda_\cft^{2d-3}}(p_H \cdot p_L),
\eeq
yielding
\beq
n_H n_L \langle \Gamma (H + L \to \cft) E\rangle = \frac{2^{2d}}{128\pi^2}\frac{ A_{d-1/2} \sec^2(\pi d)}{\Gamma(-1/2-d)\Gamma(3/2-d)}  \frac{\lambda^2_\cft}{\Lambda_\cft^{2d-3}} T^{2d+2}.
\label{eq:HLOs}
\eeq
From \Eq{eq:gamD}, we see that the scattering process is UV-dominant for $d > 2$ and IR-dominant for a narrow range of $3/2 < d < 2$.

Next, the spin-averaged amplitude for the Higgs decay, $h \rightarrow \cft + \nu$, is computed to be
\beq
\sum |\mathcal M_h|^2 = \frac{\lambda^2_\cft}{\Lambda_\cft^{2d-3}} m_h^2,
\eeq
and, in the relativistic approximation $E_h \simeq p_h$, the energy transfer rate is given by
\beq
n_h \langle \Gamma (h \rightarrow \cft + \nu) E\rangle = \frac{1}{64\pi^4} \frac{A_{d-1/2}}{(d-3/2)(d+1/2)} \frac{\lambda_\cft^2}{\Lambda_\cft^{2d - 3}} m_h^{2d-1} T^3.
\label{eq:efhcft}
\eeq
Finally, the matrix element for the gauge boson decay process, $W/Z \rightarrow \cft + \ell/\nu$, at the leading order in $\lambda_\cft$ is given by~\cite{Chacko:2020zze}
\beq
i\mathcal{M}_{W/Z} = \frac{ig_{W/Z}}{2\sqrt2} \bar u(p_\cft) \left(\frac{-i \lambda_\cft v}{\Lambda_\cft^{d-3/2}} \right) \frac{i\slashed{p}_\cft}{p_\cft^2} \gamma^\mu (1 - \gamma^5) v(p_{\ell/\nu}) \epsilon_\mu (p_{W/Z}).
\eeq
In the relativistic approximation again, we obtain
\beq
n_{W/Z} \langle \Gamma (W/Z \rightarrow \cft + \ell/\nu) E \rangle = {\mathcal N} \,\frac{g_{W/Z}^2}{48\pi^3} m^2_{W/Z}T^3,
\label{eq:efwzcft}
\eeq
where ${\mathcal N}$ is the rate of energy transfer into CFT, normalized to the rate associated with a purely SM process $n_{W/Z} \langle \Gamma (W/Z \rightarrow \nu + \ell) E_\nu \rangle$. The normalized rate is given by
\beq
\mathcal N = \frac{\lambda_\cft^2 v^2}{M_\gap^2} A_{d-1/2} \left( \frac{M_\gap}{\Lambda_\cft} \right)^{2d-3} \left[ \frac{ \sec(\pi d)}{2}
+  \frac{ (m_{W/Z}/M_\gap)^{2d-5} }{\pi(d-1/2)(d-5/2)} \mathcal R(d, M_\gap) \right].
\eeq
Here $\mathcal R(d,M_\gap)$ is a hypergeometric function 
\beq
\mathcal R(d,M_\gap) = {}_2F_1\left(\frac52-d,\frac52-d;\frac72-d;(-1)^{d-3/2}\left( \frac{d-1/2}{2} \right)^{\frac{1}{d-5/2}} \left( \frac{M_\gap}{m_{W/Z}} \right)^2 \right)
\eeq
which regulates the poles in $\sec(\pi d)$ at $d = 5/2, 7/2, \cdots$, so that our expression for the CFT production rate is well-defined for all values of $d \geq 3/2$ (the pole at $d=3/2$ is taken care of by $A_{d-1/2}$). Up to order-one factors, the normalized rate can be approximated as
\beq
\mathcal N \,\sim\, \frac{\lambda_\cft^2 v^2}{M^2_\gap}\,\left(\frac{M_\gap}{\Lambda_\cft}\right)^{2d-3},~~~3/2\leq d<5/2
\label{eq:N}
\eeq  
and
\beq
\mathcal N \,\sim\, \lambda_\cft^2 \,\left(\frac{v}{\Lambda_\cft}\right)^{2d-3},~~~ d\geq 5/2.
\eeq  
For $d < 5/2$, the energy transfer rate due to $W/Z$ decays is parametrically larger than that due to Higgs decays, by a factor of $\sim (v/M_\gap)^{5-2d}$. This is due to the enhanced effective coupling of the $W$ to the low-mass CFT states.\footnote{The IR-enhancement observed in decay is absent in the scattering process $\ell+W\to$~CFT, despite the matrix element having the same form. Momentum conservation fixes the mass of the CFT state produced in this process to be equal to the center-of-mass energy of the colliding particles, and hence the low-mass CFT states are not produced at high temperatures.} On the other hand, for $d\geq 5/2$, the energy transfer rates due to the Higgs and $W/Z$ decays are comparable. 

The expression~\Eq{eq:N} is divergent as $M_\gap\to 0$, leading one to doubt its validity in this limit. In practice, this is never an issue for the analysis of this paper. The reason is that, as will be shown in Section~\ref{sec:nu_mass}, in our model we always have
\beq
\mathcal N \,\lsim\, \frac{m^2_\nu}{M^2_\gap}\,,
 \eeq  
where $m_\nu \sim 0.1$~eV is the active neutrino mass, and phenomenological constraints on dark matter mass ensure that $\mathcal N \ll 1$ in any viable model. Nevertheless, before we proceed, let us briefly comment on this theoretical issue. The $M_\gap\to 0$ divergence is an artifact of describing the $\nu$-CFT mixing by a single insertion of the mixing operator, as in Fig.~\ref{fig:fyn}. This is exactly what happens in an ordinary field theory where ${\cal O}_\cft$ is replaced with an elementary sterile neutrino field $N$. The approximation of Fig.~\ref{fig:fyn} contains a spurious divergence when $m_N\to 0$, which disappears upon consistently resumming the diagrams with multiple mixing insertions and identifying the on-shell mass eigenstates as the out-states of the $S$ matrix. 
When this is done, the decay rate $\Gamma(W\to N + \ell)$ is simply equal to the SM rate $\Gamma(W\to \nu + \ell)$ multiplied by $\sin^2\theta$, where $\theta$ is the mixing angle between the active and sterile neutrinos. We expect that a resummation of the diagrams with multiple mixing insertions, together with a proper treatment of the external states that contain an admixture of the CFT continuum, would likewise regulate the apparent $M_\gap\to 0$ divergence in our model. Unfortunately, at the moment we are not able to demonstrate this explicitly, because we do not have a formalism to describe the mass-mixing between the elementary active neutrino and the continuum of CFT states beyond the leading mass-insertion approximation. 

\subsection{Relic Density}

The CFT energy density as a function of the Standard Model bath temperature $T$ can be obtained by integrating the Boltzmann equation given in \Eq{eq:btzm}. Since the scattering and decay processes exhibit different temperature dependencies, we first integrate the Boltzmann equation for the scattering case \Eq{eq:HLOs}. In this case, we obtain
\beq
\rho_{\cft}^s(T) = \frac{3\sqrt5 M_{pl}}{4(2-d)\pi^{\frac{3}{2}}\sqrt{g_*(T)}}B_\Gamma T^4 \left(T_R^{2d-4} - T^{2d-4}\right),
\eeq
where
\beq
B_\Gamma \equiv \frac{2^{2d}}{128\pi^2}\frac{ A_{d-1/2} \sec^2(\pi d)}{\Gamma(-1/2-d)\Gamma(3/2-d)}  \frac{\lambda^2_\cft}{\Lambda_\cft^{2d-3}}.
\eeq
As anticipated above, the result is approximately insensitive of $T_R$ in the IR-dominated regime $3/2<d<2$, and is strongly $T_R$-dependent in the UV-dominated regime $d>2$. 

Next we consider the decay processes \Eqs{eq:efhcft}{eq:efwzcft}.
The calculation can be performed analytically in the relativistic approximation where the Higgs and $W/Z$ are assumed to be described by a massless Maxwell-Boltzmann distribution. The decay processes become important around the electroweak scale, $T \sim v$. Consequently, we obtain
\beq
\rho^d_\cft(T) = \frac{\sqrt5 M_{pl}}{2\pi^{\frac{3}{2}}\sqrt{g_*(T)}}C_\Gamma \frac{T^4}{v} \left(\frac{v^3}{T^3} - 1\right),
\eeq
where $v$ is the electroweak VEV and 
\beq
C_\Gamma \equiv \frac{g_{W/Z}^2}{48\pi^3} \mathcal N + \frac{1}{64\pi^4} \frac{A_{d-1/2}}{(d-3/2)(d+1/2)} \frac{\lambda_\cft^2}{\Lambda_\cft^{2d - 3}} \frac{m_h^{2d-1}}{v^2}
\eeq
is the ratio of the energy transfer rates from $W$ and $h$ decays, which is independent of temperature.

Around $T\sim m_*$, where $m_*\approx 100$~GeV is the mass scale of the Higgs and the weak gauge bosons, the initial state Higgs or weak gauge bosons, relevant for both decay and scatterings, falls out of the thermal bath, and the corresponding energy transfer is effectively terminated. The energy density present in the dark sector is then simply given by
\begin{equation}
\begin{aligned}
\rho_\cft(m_*) &= \rho^s_\cft(m_*) + \rho^d_\cft(m_*) \\
&= \frac{\sqrt5 M_{pl}}{2\pi^{\frac{3}{2}}\sqrt{g_*(m_*)}}  
\left[ \frac{3}{2(2-d)} B_\Gamma m_*^4 \left(T_R^{2d-4} - m_*^{2d-4}\right)
+ C_\Gamma \frac{m_*^4}{v} \left(\frac{v^3}{m_*^3} - 1\right) \right].
\end{aligned}
\end{equation}
The following features of this result are important to keep in mind:

\begin{itemize}

\item For $3/2\leq d < 2$, the CFT energy density is dominated by the $W/Z$ decays, and is independent of the reheat temperature ({\it i.e.} IR-dominant freeze-in);

\item For $2\leq d < 5/2$, the CFT energy density is dominated by the $W/Z$ decays, and is independent of the reheat temperature, as long as $T_R$ is sufficiently low; specifically,
\beq
T_R \lsim m_* \Bigl( \frac{v}{M_\gap}\Bigr)^{\frac{1}{2d-4}-1}.
\eeq
For larger $T_R$, the energy density is dominated by the scattering process, and depends sensitively on the reheat temperature;

\item For $d>5/2$, the energy density is dominated by the scattering process, and depends sensitively on the reheat temperature ({\it i.e.} UV-dominant freeze-in).

\end{itemize}

Below $T \sim m_*$, $\rho_\cft$ redshifts as radiation until the dark sector's temperature $T_D$ approaches the mass of the dark matter candidate.\footnote{Before the phase transition in the dark sector, this scaling is a consequence of conformal symmetry. After the phase transition, we generically expect that the dark sector will be radiation-dominated for $T_D\gsim m_\chi$, but the actual scaling 
may be more complicated depending on the details of the bound state masses and decay rates.} Once $T_D$ drops below the dark matter mass, it quickly becomes non-relativistic and its energy density redshifts as matter ($\rho_\cft \propto a^{-3}$). Hence the dark matter relic density today is given by
\beq
\rho_\cft (T_0) = \frac{g_*(T_0)}{g_*(m_*)} \left( \frac{T_0}{T_m} \right)^3  \left( \frac{T_m}{m_*} \right)^4 \rho_\cft(m_*),
\eeq
where $T_m$ is the SM temperature at which $T_D$ drops to $m_\chi$. $T_m$ can be found from the following relation: 
\beq
\frac{Am_\chi^4}{\rho_\sm(T_m)} = \left( \frac{g_*(T_m)}{g_*(m_*)} \right)^{1/3} \frac{\rho_\cft(m_*)}{\rho_\sm(m_*)},
\eeq
where $A$ is equal to the number of degrees of freedom in the dark-sector CFT (up to an order-one factor). The interesting parameter range in our model corresponds to $T_m$ around or somewhat below the QCD scale, so that $g_*(T_m) \sim \mathcal O(10)$.

\begin{figure}[t!]
\centering
      \includegraphics[width=0.49\linewidth]{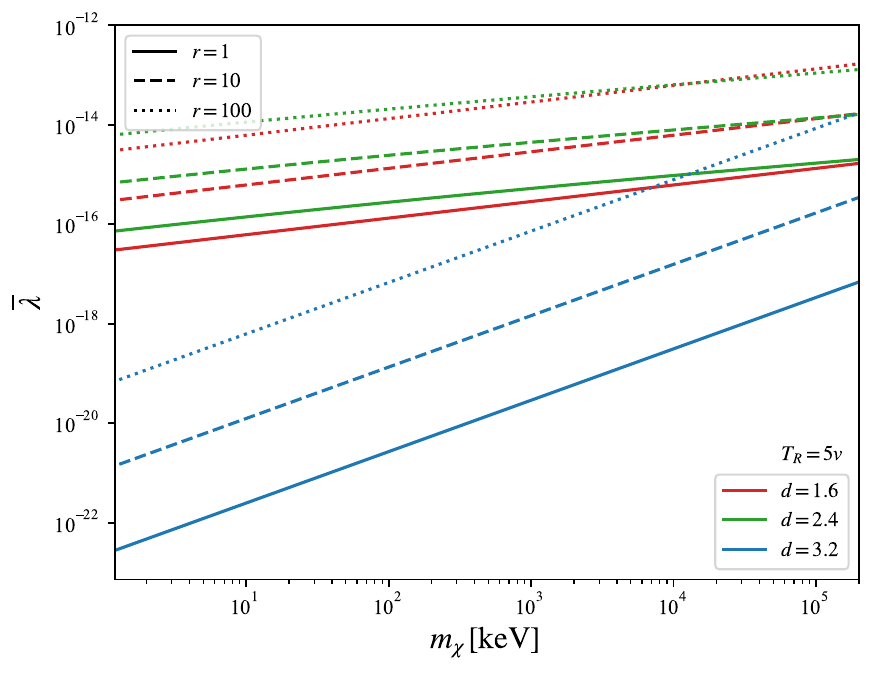}
    \includegraphics[width=0.49\linewidth]{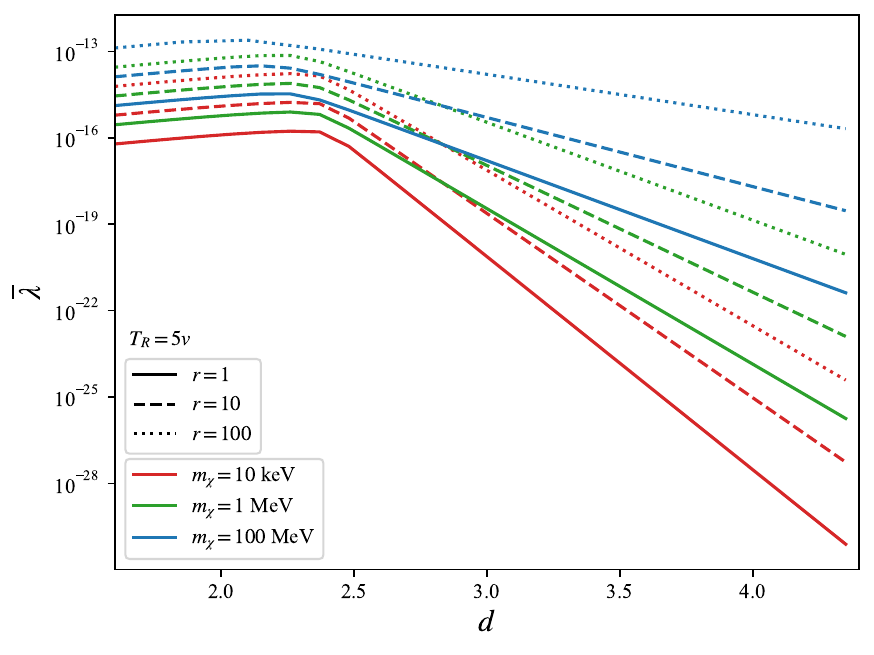}   
\caption{The effective CFT coupling $\overline \lambda \equiv \lambda_\cft \left( \frac{\mu_\text{IR}}{\Lambda_\cft} \right)^{d-3/2}$ that reproduces the observed relic density w.r.t the DM mass $m_\chi$ (left) and dimension $d$ (right), for reheating temperature $T_R = 5 v$. Solid, dashed and dotted lines represent $r = \frac{M_\gap}{m_\chi} = 1, 10, 100$ respectively. In the left panel, the parameter $d$ is varied as $1.6$, $2.4$, and $3.2$, represented by red, green, and blue lines, respectively. In the right panel, $m_\chi$ is varied as 10 keV, 1 MeV, and 100 MeV, also shown by red, green, and blue lines, respectively. }
\label{fig:cofi_lam}
\end{figure}

In Figure~\ref{fig:cofi_lam} we show the effective COFI coupling
\beq
\overline \lambda \equiv \lambda_\cft \left( \frac{M_\gap}{\Lambda_\cft} \right)^{d-3/2}
\eeq
that reproduces the correct relic density. In the left panel, the effective coupling is plotted against the DM mass $m_\chi$, while in the right panel, it is plotted against the scaling dimension $d$. In both cases, we choose reheat temperature to be $T_R=5v\sim 1.2$~TeV. As discussed above, for $d<2.5$, the relic density is approximately independent of $T_R$, while for $d>2.5$ increasing $T_R$ would decrease the effective coupling corresponding to correct relic density. Solid, dashed and dotted lines corresponds to $r =  \frac{M_\gap}{m_\chi} = 1, 10, 100$ respectively. (Note that a hierarchy $m_\chi \ll M_\gap$ is technically natural for both fermionic and PGB DM candidates.)
In the left panel, the values of $d$ are set to 1.6, 2.4, and 3.2, indicated by red, green, and blue lines, respectively. In the right panel, $m_
\chi$   takes values of 10 keV, 1 MeV, and 100 MeV, also represented by red, green, and blue lines. As typical in freeze-in scenarios, correct DM abundance requires a highly suppressed effective coupling between the SM and the dark sector. Such small couplings are in fact very natural in the model we consider: for example, in a model with $m_\chi=10$~keV, $M_\gap=1$~MeV, $d=2.4$, $\lambda_\cft=0.1$, and $\Lambda_\cft=2.5\times 10^{12}$~GeV, the fermionic COFI mechanism produces the DM abundance approximately consistent with observations. Alternatively, a model with $\lambda_\cft \ll 1$ and $\Lambda_\cft \sim \mathcal{O}$~(TeV) can be achieved in a natural way~\cite{Hong:2019nwd,Hong:2022gzo}. Such a possibility is realized by starting in the deep UV with a coupling between the SM and a gauge theory sector. The latter gauge theory sector flows to a strongly coupled IR fixed point, generating the COFI interaction. The point is that the interaction between the SM and the gauge theory sector is generically irrelevant, and thus under the renormalization group flow is reduced significantly without any input small parameters. See \cite{Hong:2019nwd,Hong:2022gzo} for more details.

In this section, we have determined the portal interaction strength needed to produce the correct amount of DM. Can this portal interaction also provide sufficient breaking of the CFT to give the DM its mass? As discussed in Section~\ref{sec:fCOFI}, for $d\geq 2$ the answer is negative, since the deformation induced by the portal is negligible in the IR. For $3/2\leq d <2$, an IR CFT breaking is generated. The breaking scale is given by \Eq{eq:hlogap}, which can be rewritten as 
\beq
M^{(HL)}_\gap \sim \frac{\overline \lambda^2 \Lambda_\sm}{16 \pi^2}.
\eeq
Even with the highest plausible value of the cut-off, $\Lambda_\sm\sim M_{pl}$, this scale remains well below keV throughout the parameter space consistent with the DM relic density. Since the DM mass is required to be above a few keV by the Lyman-$\alpha$ constraints, the CFT breaking generated by the portal interaction alone is not sufficient. We conclude that for any value of $d$, an additional source of CFT breaking must be introduced, as was done in \Eq{eq:extra}. 


\section{Neutrino Mass Generation}
\label{sec:nu_mass}

The CFT operator $\mathcal O_\cft$, and the composite states created by it in the hadronic phase of the dark CFT, carry the quantum numbers of a right-handed neutrino. Thus, we expect the portal interaction of~\Eq{eq:hln} to generate a mass for active neutrinos. In this section, we will show that in certain dark-sector models, this mass could account for the observed neutrino mass scale $m_\nu \sim 0.1$~eV, leading to a very attractive possibility that the same dynamics is responsible for the observed dark matter abundance and neutrino mass. This will be discussed in subsection~\ref{sec:Dirac}. An alternative (perhaps more generic) possibility is that the neutrino mass generation is achieved by a separate mechanics disconnected from the DM sector. The simplest realization of this will be outlined in subsection~\ref{sec:RHN}.  

\subsection{Neutrino Mass from the Dark Matter Portal}
\label{sec:Dirac}

The CFT operator interpolation~\Eq{eq:cftint} and the electroweak symmetry breaking give rise to a possible mass source for active neutrinos:
\beq
\frac{\lambda_\cft}{\Lambda_\cft^{d - 3/2}} (HL) \mathcal O_\cft \sim \lambda_\cft v \sum_n a_n \left( \frac{M_\gap}{\Lambda_\cft} \right)^{d - 3/2} (\nu_L \psi_n)\,=\,\sum_n a_n \overline{\lambda} v (\nu_L \psi_n),
\label{banor}
\eeq
where $\psi_n$ are right-handed fermions with the quantum numbers of a sterile neutrino, composed of confined CFT degrees of freedom. Generically, we expect that these fermions get masses of order $M_\gap$ from confining dynamics. These masses may be Majorana or Dirac. In the Majorana case, the mainly-active neutrino gets a mass
\beq
m_\nu \sim \frac{\overline{\lambda}^2 v^2}{M_\gap}\,,
\eeq
where we assumed $a_n\sim 1$. Unfortunately, for the values of parameters required to obtain the correct DM relic density, shown in~\Fig{fig:cofi_lam}, this is much smaller than the observed neutrino mass scale, $\sim 0.1$~eV. Alternatively, if we choose COFI parameters to obtained correct neutrino masses, it will lead to overclosure of the universe. Either way, this particular mechanism for the neutrino mass generation can not accommodate both requirements. 

\begin{figure}[t!]
	\centering
	\includegraphics[width=0.49\linewidth]{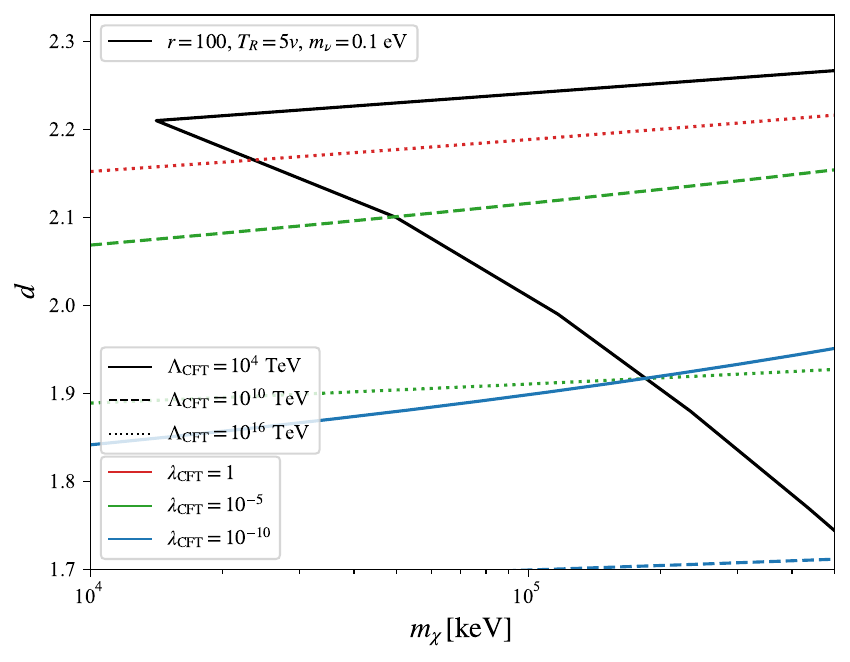}
	\includegraphics[width=0.49\linewidth]{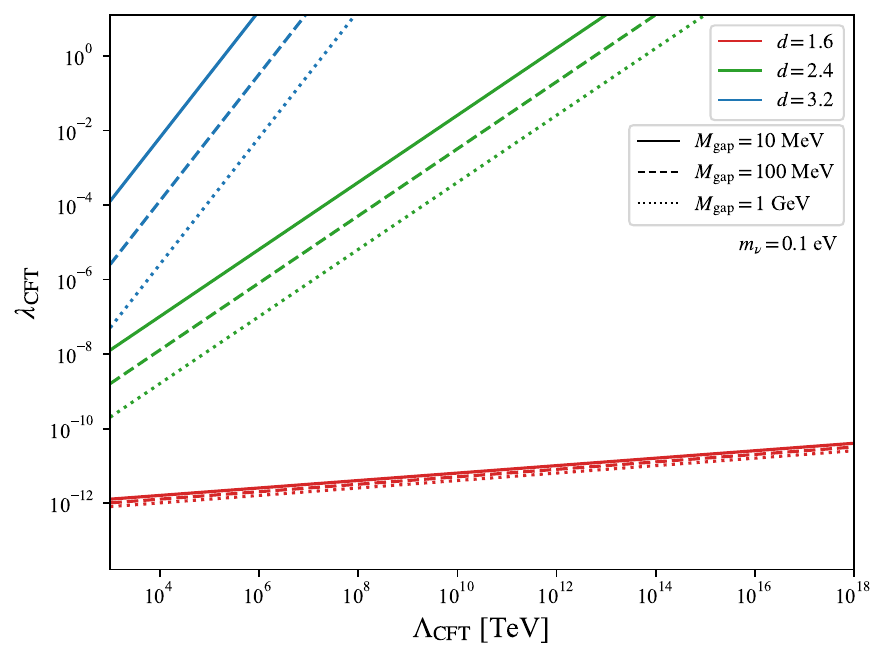}
	\caption{\textbf{Left}: Red, green, and blue curves show the parameters corresponding to the measured neutrino mass scale, $m_\nu=0.1$~eV, for several representative values of $\lambda_\cft$ and $\Lambda_\cft$. The black curve shows the parameters where the correct DM relic density is reproduced. It is assumed that $r = \frac{M_\gap}{m_\chi} = 100$ and $T_R = 5v$.   \textbf{Right}: The relation between $\lambda_\cft$ and $\Lambda_\cft$ that generate the correct neutrino mass $m_\nu = 0.1 \ev$, for a few representative values of $d$ and $M_\gap$.}
	\label{fig:cofi_dir}
\end{figure}

In the Dirac case, $\psi_n$ pair up with left-handed composite fermions $\psi^\prime_n$. This will be the case if the dark CFT respects the lepton number and hence no Majorana mass is allowed. In this case, the composite fermion spectrum will be purely vector-like, and the addition of the field $\nu_L$ leads to a mismatch between left- and right-handed states, so that there is always one massless left-handed fermion left in the spectrum even after including~\Eq{banor}. Thus, active neutrinos, $\nu_L$, are remained to be massless.\footnote{Alternatively, this may be seen from chiral symmetry of the theory. To simplify the discussion, let us consider the truncated picture where the confined phase of the dark sector is represented by a singe lightest Dirac composite fermion. Then, the fermion system is described by $\mathcal{L} \sim m \nu_L \psi_1 + M \chi_1 \psi_1$. This theory has $U(1)_L^2 \times U(1)_R$ chiral symmetry, two and only two of which are broken by mass terms. Therefore, there exists one chiral symmetry unbroken, hence one massless fermion. This argument can easily be extended to a tower of Dirac fermion coupled to a chiral fermion, and details can be found in \cite{Hong:2019bki}.} To avoid this conclusion, the composite dynamics must produce a {\it chiral} spectrum. Specifically, a massless composite right-handed fermion $\psi_0$, with no left-handed composite partner, is required. Such composite chiral massless fields may arise naturally in theories where they are required by 't Hooft anomaly matching.\footnote{See~\cite{Weinberg:1996kr} for standard discussion of 't Hooft anomaly matching. For a 't Hooft anomaly associated with continuous symmetries, there exist a theorem by Coleman and Grossman~\cite{Coleman:1982yg}  based on S-matrix technique which states that such a anomaly has to be matched by massless state with helicity $\pm 1/2$. A modern discussion on 't Hooft anomalies associated with discrete as well as continuous symmetries, including those of generalized symmetries (see \cite{Schafer-Nameki:2023jdn, Brennan:2023mmt, Bhardwaj:2023kri, Shao:2023gho} for reviews), and various ways they can be matched, can be found for e.g.~in section 3 of \cite{Anber:2021iip}.} If such a massless composite fermion $\psi_0$ is present, the portal interaction pairs it up with $\nu_L$ to produce an active neutrino with a Dirac mass,   
\beq
m_\nu \sim \overline{\lambda} v\,,
\label{eq:mnuso}
\eeq
which may be of order 0.1~eV provided that
\beq
\overline{\lambda}\sim 10^{-13}.
\eeq
Such values of effective coupling can be consistent with the observed DM relic density as well; see Fig.~\ref{fig:cofi_dir}.

The existence of a very light composite fermion $\psi_0$ implies that the dark matter particle $\chi$ is no longer the lightest dark-sector composite state. A phenomenologically viable COFI model requires that the energy stored in the dark sector ends up in the form of DM particles today. But composite states generically interact strongly with each other, leading to reactions such as $\chi\chi\to\psi_0\psi_0$ which would transfer most of the energy to $\psi_0$ at late times. To achieve a viable model, these reactions must be strongly suppressed. This requires additional symmetry structure in the hadronic phase of the dark sector: either $\chi$ and/or $\psi_0$ carry conserved quantum numbers that forbid the decay and annihilation reactions, or perhaps there is an initial asymmetry in a quantum number carried by $\chi$ and not $\psi_0$. Note that this requirement essentially precludes the possibility that $\chi$ itself is a sterile neutrino state, interpolated by $\mathcal O_\cft$. However, other possibilities, including PGB dark matter, are still open. Detailed model-building along these lines is beyond the scope of this paper and is left for future work.    

Using \Eq{eq:mnuso}, the COFI production rates in \Eq{eq:HLOs}, (\ref{eq:efhcft}) and (\ref{eq:efwzcft}) can be expressed in terms of $M_\gap$ and $m_\nu$: 
\bea
&& \hspace{-0.8cm} n_H n_L \langle \Gamma (H + L \to + \cft) E\rangle = \frac{2^{2d}}{128\pi^2}\frac{ A_{d-1/2} \sec^2(\pi d)}{\Gamma(-1/2-d)\Gamma(3/2-d)} \frac{m_\nu^2}{v^2} M_\gap^{3-2d} T^{2d+2} \\
&& \hspace{-0.8cm} n_h \langle \Gamma (h \rightarrow \cft + \nu) \rangle =  \frac{1}{64\pi^4} \frac{A_{d-1/2}}{(d-3/2)(d+1/2)}  \frac{m_\nu^2}{v^2} M_\gap^{3-2d} m_h^{2d-1} T^3 \label{eq:geso} \\
&& \hspace{-0.8cm} n_{W/Z} \langle \Gamma (W/Z \rightarrow \cft + \ell/\nu) E \rangle = \mathcal N\,\frac{g_{W/Z}^2}{48\pi^3} \, m^2_{W/Z}T^3,
\eea
with
\beq
\mathcal N =\frac{m_\nu^2}{M_\gap^2} A_{d-1/2} \left[ \frac{ \sec(\pi d)}{2}   + \frac{ (m_{W/Z}/M_\gap)^{2d-5} }{\pi(d-1/2)(d-5/2)} \mathcal R(d, M_\gap) \right],
\eeq
where once again we substituted $\mu_\ir = M_\gap$ explicitly. Up to order-one factors, we find
\beq
\mathcal N \,\sim\, \frac{m_\nu^2}{M^2_\gap}\,\ll 1,~~~3/2\leq d<5/2,
\eeq  
justifying the use of a single mass-mixing insertion approximation made in Section~\ref{sec:fCOFI}. 

The left panel of Figure~\ref{fig:cofi_dir} shows the relation between $m_\chi$ and $d$ required to yield the correct DM relic abundance and neutrino mass $m_\nu = 0.1 \ev$, for a few representative values of $\lambda_\cft$ and $\Lambda_\cft$. 
Here we assume that all energy transferred to the dark sector is stored in the form dark matter $\chi$, with no leakage into neutrinos via the light $\psi_0$ state. It is assumed that $T_R = 5v$. The relic density curve exhibits a sharp turn starting from $d\gtrsim 2.2$. This is because at this point the UV-dominant scattering process becomes the dominant source of DM production, bringing in strong dependence on the reheating temperature. In the right panel of Figure~\ref{fig:cofi_dir}, we also analyze the relation between $\lambda_\cft$ and $\Lambda_\cft$ that produces a neutrino mass $m_\nu = 0.1 \ev$ with the choices of $d$ and $M_\gap$.
We conclude that the fermionic COFI scenario can account for the observed dark matter and neutrino masses simultaneously, generating both from the same portal interaction between the SM and the dark sector. Phenomenological constraints on this scenario will be discussed in Section~\ref{sec:pheno}. 

\subsection{Elementary Right-handed Neutrino}
\label{sec:RHN}

The possibility that the DM and neutrino masses arise from a common mechanism is fascinating, but it is of course also possible that the two phenomena have nothing to do with each other. In particular, while the portal interaction postulated in the fermionic COFI scenario will always make a non-zero contribution to active neutrino masses, this may be subdominant to other contributions. To construct an explicit model realizing this, consider a dark sector with vector-like composite fermions with masses of order $M_\gap$, and add an {\it elementary} right-handed neutrino field $\nu_R$. The Yukawa coupling between the lepton doublet $L$ and $\nu_R$ is the dominant contribution to the active neutrino mass. (As usual, the smallness of $m_\nu$ compared to the weak scale can be due to either the see-saw mechanism or a very small value of the Yukawa coupling.) In addition, the COFI portal interaction induces mass mixing between $\nu_L \in L$ and the tower of composite states:
\beq
\mathcal L_\mathrm{IR} \supset \lambda_\cft v \sum_n a_n \left( \frac{M_\gap}{\Lambda_\cft} \right)^{d - 3/2} (\nu_L \psi_n).
\eeq
We assume that
\beq
m_\nu \gg \lambda_\cft \,v \, \left( \frac{M_\gap}{\Lambda_\cft} \right)^{d - 3/2},
\eeq
so that the effect of the COFI portal on the active neutrino mass is subdominant. 
Incidentally, this imples an upper bound on the normalized rate of $W/Z$ decays introduced in Section~\ref{sec:fCOFI}:
\beq
\mathcal N \ll \frac{m^2_\nu}{M^2_\gap} \ll 1\,,
\eeq
so the approximation of single mass-mixing insertion used in our relic density calculations is justified in this model as well. 

The COFI portal generates a mixing between the active and (composite) sterile neutrinos:
\beq
|\theta_n| \simeq \frac{\lambda_\cft a_n v}{m_{\psi_n}} \left(\frac{M_\gap}{\Lambda_\cft} \right)^{d-3/2}.
\eeq
If the lightest composite state $\psi_0$ is identified as the dark matter candidate, this mixing leads to the possibility of a decay $\psi_0\to \nu \gamma$, providing a promising potential signature (and a relevant phenomenological constraint) of this model. This will be discussed further in Section~\ref{ssec:ddm}.

\section{Dark Matter Phenomenology}
\label{sec:pheno}

In freeze-in scenarios such as COFI, the interaction between DM and SM particles are extremely weak, rendering many of the usual dark matter constraints -- such as those from direct, indirect, and collider searches -- largely ineffective. However, dark matter self-interactions, large-scale structure, and stellar cooling rates can still provide relevant constraints. The self-interactions and large-scale structure are independent of the DM-SM coupling, while in stellar cooling, the small coupling is offset by the large number of SM particles within stellar bodies.
Moreover, unlike PGB dark matter, sterile neutrino dark matter is unstable and produces potentially observable X-ray signals due to its mixing with active neutrinos. 

\subsection{DM Self-Interaction}

For a generic bound state within a strongly-coupled theory with the mass of order $M_\gap$, the elastic self-scattering cross section can be estimated as the geometric cross section, given by
\beq
\sigma^\mathrm{SI} \sim \frac{1}{8\pi M_\gap^2}.
\eeq
DM self-interaction cross section is constrained by observations of galaxy cluster dynamics ({\it e.g.} the Bullet cluster): the current bound is $\sigma^\mathrm{SI}/m_\chi \lesssim 4500 \gev^{-3}$~\cite{Markevitch:2003at,Randall:2008ppe,Robertson:2016xjh}. For composite DM with mass at the gap scale, this implies 
\beq
m_\chi \gsim 20~\mev.
\eeq
In this mass range, correct DM relic density can be generated, but the required values of $\tilde{\lambda}\sim 10^{-18}$ are too small to also generate the neutrino mass via the COFI portal operator. Viable models can still be constructed by adding an elementary right-handed neutrino; see Sec.~\ref{sec:RHN}.     

\begin{figure}[t]
\centering
  \includegraphics[width=0.7\linewidth]{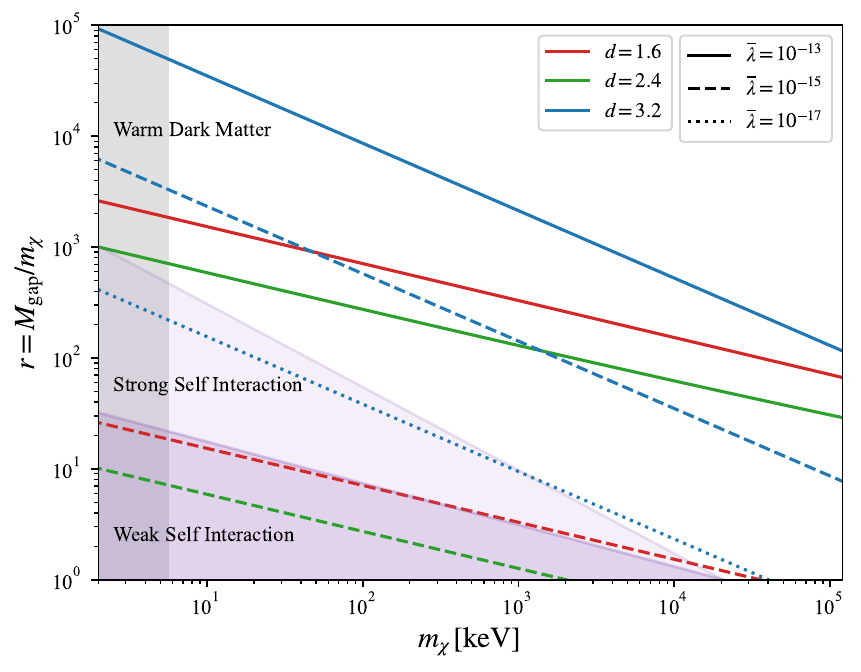}
\caption{The ratio of $M_\gap$ to $m_\chi$ that produces the observed relic density for fixed values of $\overline{\lambda}$. The ``Weak Self Interaction'' refers to \Eq{eq:sicrs1}; the ``Strong Self Interaction'' to \Eq{eq:sicrs2}. Solid, dashed and dotted lines correspond to different values of $\overline{\lambda}$, while red, green, blue curves represent different choices of $d$. }
\label{fig:ratio_d}
\end{figure}

Allowing the DM particle to be lighter than $M_\gap$ with $r = M_\gap / m_\dm \gg 1$ can weaken the self-interaction constraint. 
In this case, the DM particles are described within a weakly-coupled effective theory of the hadronic phase, and need not have geometric self-interaction cross sections. Besides dark matter $\chi$, the IR effective theory typically includes a collection of bound states with masses around $M_\gap$. These states will couple to $\chi$ and mediate DM self-interactions. Following the previous work~\cite{Hong:2019nwd,Hong:2022gzo,Chiu:2022bni}, we model the interaction for PGB dark matter as
\beq
\frac{g_*}{M_\gap} \phi (\partial\chi)^2, \qquad g_* \rho^\mu \chi^* \partial_\mu \chi
\label{eq:selfpgb}
\eeq
for a scalar mediator $\phi$ and vector mediator $\rho^\mu$ respectively. 
Similarly, the leading couplings of fermionic dark matter to a pseudo-Goldstone boson $\pi$, scalar $\phi$, and vector $\rho^\mu$ are as follows:
\beq
\frac{g_*}{M_\gap} \partial_\mu \pi \chi^\dagger \bar\sigma^\mu \chi,
\qquad
g_* \phi \bar\chi \chi,
\qquad
g_* \rho^\mu \chi^\dagger \bar\sigma^\mu \chi.
\label{eq:selfsn}
\eeq
The characteristic dimensionless coupling is estimated to be $g_* \sim \frac{4\pi}{\sqrt{N}}$ in the large-$N$ limit.; we will assume $g_*\sim 1$ in the estimates below.

The self-interaction cross-section is given by
\beq
\sigma^\mathrm{SI}_\chi \sim \frac{1}{8\pi r^6 M^2_\gap} \hspace{0.8cm} [{\rm Weak \;\; Self \;\; Interaction}]
\label{eq:sicrs1}
\eeq
for PGB dark matter with a scalar mediator and fermionic DM with a pseudoscalar mediator. We will denote this as ``Weak Self Interaction''. For the other three possibilities mentioned above, we have
\beq
\sigma^\mathrm{SI}_\chi \sim \frac{1}{8\pi r^2 M^2_\gap} \hspace{0.8cm} [{\rm Strong \;\; Self \;\; Interaction}]
\label{eq:sicrs2}
\eeq
which we will dub ``Strong Self Interaction''.
The allowed parameter space is illustrated in Figure~\ref{fig:ratio_d}. With the additional suppression of the self-interaction cross section,  DM masses as low as 10 keV are allowed. Values of $r$ of order $10 - 100$ are sufficient to evade observational constraints in models with stronger suppression, \Eq{eq:sicrs1}. Models with weaker suppression, \Eq{eq:sicrs1}, require $r \gtrsim 10^2$. With $r$ in the $10^{3}-10^4$ range, models with $\overline{\lambda}\sim 10^{-13}$, where the neutrino mass can be generated by the COFI portal operator, are phenomenologically viable. Note that $r$ is a free parameter and small $1/r$ is radiatively stable in both the PGB and fermionic DM models due to the associated global symmetries protecting the small masses.

\subsection{Warm Dark Matter}

Dark matter production in the COFI scenario is non-thermal, in the sense that the dark sector never comes into thermal equilibrium with the SM. However, strong interactions in the dark sector itself produce a thermal distribution of energies within that sector, albeit at a temperature lower than that of the SM. This means that after the phase transition in the dark sector is completed and dark matter particles are formed, they move with near-relativistic velocities. On the other hand, since DM is free-streaming, large scale structure constraints require it to be non-relativistic during the structure formation. By leveraging the strong correlation between the local Lyman-$\alpha$ optical depth and the local DM density, observations of absorption lines from distant quasars can put an upper bound on the mass of DM. The most stringent bound arises from free-streaming requirements based on the Lyman-$\alpha$ matter power spectrum at red-shifts of $z \sim (4.2 - 5)$, which can establish a lower bound of $m_\chi \gtrsim 5.7 \kev$ at 95\% C.L. \cite{Irsic:2023equ}. This bound applies in the COFI scenario, and is shown in Fig.~\ref{fig:ratio_d}.

\subsection{Stellar Cooling}

Dark matter in the few keV -- MeV mass range can be thermally produced in stars, potentially contributing to additional cooling. However, the temperatures within most types of stellar objects, such as main sequence stars (e.g., the Sun), red giants, and horizontal branch stars, are of order of keV~\cite{Hardy:2016kme}, which falls below the mass of such dark matter candidates. Hence, there is no significant bound on our scenario from these objects.

In the core of an exploding supernova, such as the SN1987A, temperatures of order 30 MeV may be reached, so that production of DM states is kinematically allowed. Supernova constraints in the case of ordinary sterile neutrinos are well-studied in the literature~\cite{ Shi:1993ee, Bolton:2019pcu}. The dominant production channel for sterile neutrino is via vacuum oscillations of active neutrinos. Oscillation probability is dominated by the states nearest in mass to the active neutrino. In the case of COFI models, these states are the composite bound states formed after confinement, which behave essentially like sterile neutrinos. Mixing of active neutrinos with heavier composite fermion states are suppressed due to their larger masses. Therefore, in Figure~\ref{fig:sn_lambda}, we present the supernova constraints on ordinary sterile neutrinos to provide a point of reference for sterile neutrino-like DM. For PGB DM, the lightest fermionic composite of the CFT would typically lie closer to $M_\gap$ instead of $m_\chi$, which means that in these models the supernova bound could be further weakened by a factor of up to $r^2$.

\subsection{Decaying Dark Matter}
\label{ssec:ddm}

As already mentioned in Section~\ref{ssec:DMC}, in models where the dark matter candidate is the lightest fermionic state of the CFT operator $\mathcal O_\cft$, \Eq{eq:cftint}, the DM is necessarily unstable. In this section, we will consider the phenomenological constraints (and a potential signature) that arise due to this lack of stability. It should be kept in mind that these constraints do not apply to all realizations of COFI DM: for example, a PGB dark matter candidate can be exactly stable (up to Planck-suppressed interactions) due to a global symmetry. 

\subsubsection*{Lifetime of DM}

\begin{figure}[t!]
\centering
  \includegraphics[width=0.55\linewidth]{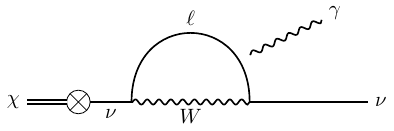}
\caption{Loop-induced radiative decay of sterile neutrino dark matter $\chi \to \nu + \gamma$.}
\label{fig:nuga}
\end{figure}

For dark matter masses below twice the electron mass, the dominant decay channel of the sterile neutrino-like DM is $\chi\to3\nu$. The total decay width, accounting for all possible combinations of neutrino flavors, is given by \cite{Pal:1981rm,Barger:1995ty}
\beq
\Gamma_{\chi\to3\nu} = \frac{G_F^2 m_\chi^5}{96\pi^3} \theta^2  \approx \frac{\theta^2}{1.5 \times 10^{14}\sec} \left( \frac{m_\chi}{10\kev} \right)^5,
\eeq
where $\theta^2$ is the total mixing angle for all lepton flavors:
\beq
\theta^2 \equiv \sum_{\alpha = e, \mu, \tau} |\theta_\alpha|^2.
\eeq
The lifetime of dark matter must be longer than the age of the Universe, $t_U = 4.4 \times 10^{17} \sec$ \cite{Planck:2018vyg}, which leads to the bound on the mixing angle \cite{Boyarsky:2018tvu}:
\beq
\theta^2 \lesssim 3.3 \times 10^{-4} \left( \frac{10 \kev}{m_\chi} \right)^5.
\eeq

\subsubsection*{X-rays}

\begin{figure}[t!]
\centering
  \includegraphics[width=0.7\linewidth]{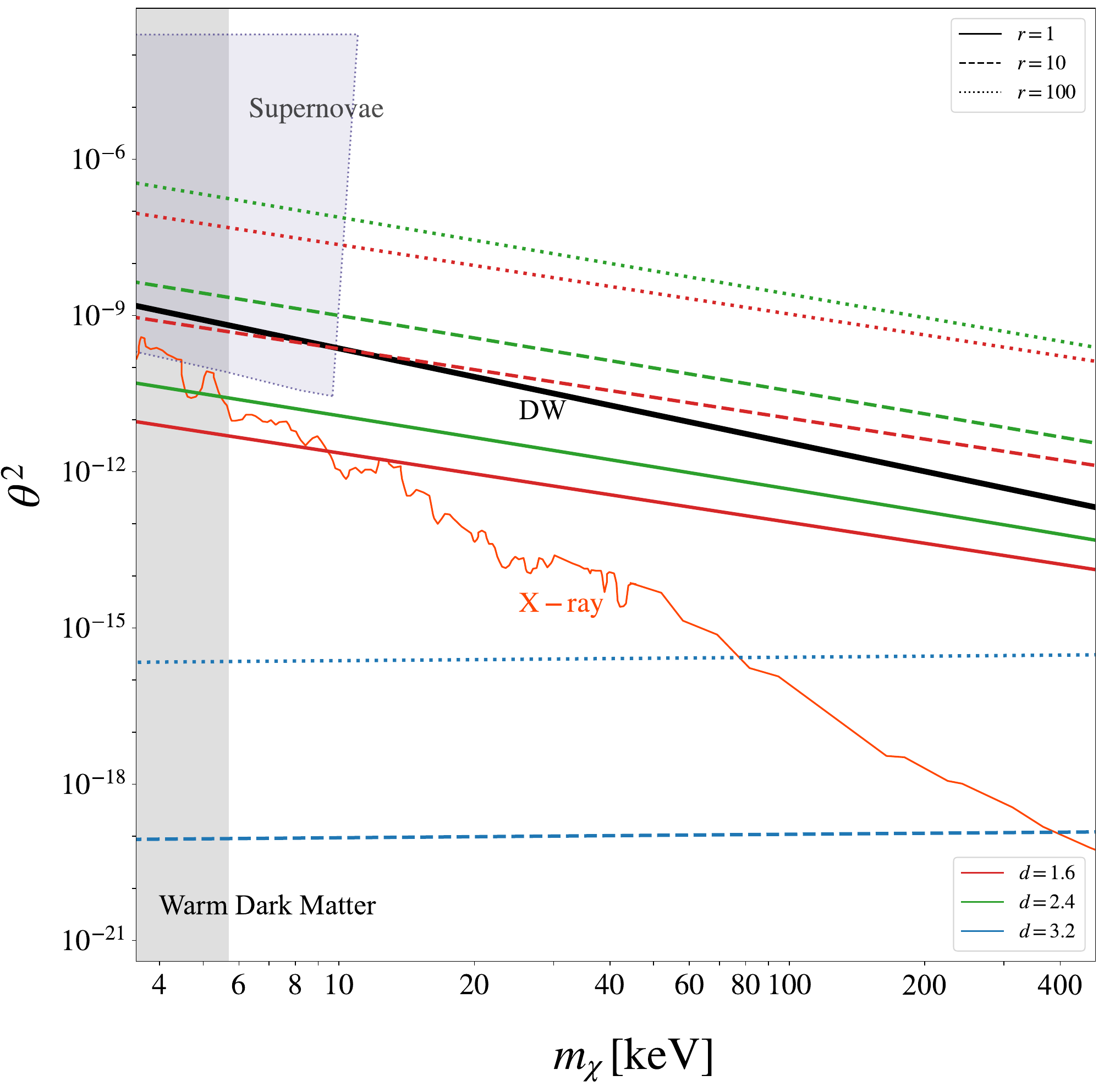}
\caption{The existing limits on the total mixing angle $\theta^2$ for sterile neutrino-like DM. Solid, dashed and dotted lines represent predictions of the fermionic COFI DM scenario with $r = \frac{M_\gap}{m_\chi} = 10, 10^2, 10^3$ respectively, while black, blue, red curves correspond to $d = 1.6, 2.4, 3.2$ respectively. The black ``DW'' line represents Dodelson-Widrow mechanism~\cite{Dodelson:1993je} for ordinary particle sterile neutrino DM.}
\label{fig:sn_lambda}
\end{figure}

A far stronger bound on the total mixing angle $\theta^2$ arises from a loop mediated radiative decay $\chi \to \nu + \gamma$, see Figure~\ref{fig:nuga}, which produces an observable photon.
The decay width of this process is given by \cite{Pal:1981rm,Barger:1995ty}
\beq
\Gamma_{\chi \to \nu + \gamma} = \frac{9\alpha G_F^2}{256\pi^2} \theta^2 m_\chi^5 \approx 5.5 \times 10^{-22} \theta^2 \left( \frac{m_\chi}{1\kev} \right)^5 \text{sec}^{-1}.
\eeq
This decay results in the emission of a photon with energy $E_\gamma = \frac{m_\chi}{2}$. Astrophysical observations, which are highly sensitive to such a monochromatic signal, place stringent constraints on the parameter space of sterile neutrino DM~\cite{Watson:2011dw,Horiuchi:2013noa,Bulbul:2014sua,Boyarsky:2014jta,Perez:2016tcq,Ng:2019gch,Dessert:2018qih,Roach:2019ctw,Foster:2021ngm,Sicilian:2022wvm}.

The current limits on the total mixing angle $\theta^2$ for sterile neutrino-like DM from supernova, Lyman-$\alpha$ data, and X-ray observations, are presented in Figure~\ref{fig:sn_lambda}. Predictions of the fermionic COFI dark matter scenario are shown by the colored lines; the colors and styles of the curves are the same as in Figure~\ref{fig:cofi_lam}. For comparison, the black line shows the prediction of the ordinary particle sterile neutrino DM model, where the DM is produced via the Dodelson–Widrow (DW)~\cite{Dodelson:1993je} mechanism. While the particle sterile neutrino DM is completely ruled out by the X-ray constraints, the sterile neutrino-like composite DM in the fermionic COFI model can be consistent with observations. In particular, models with $d > 5/2$ can easily evade all existing constraints. Such models may form an attractive target for future X-ray observations. Note that the constraints shown here are conservative, since increasing the reheating temperature or incorporating additional channels to produce more CFT states only serves to reduce $\overline{\lambda}$ and $\theta$.

\section{Conclusions}
\label{sec:con}

In this paper, we investigated a theoretical scenario of ``Conformal Freeze-In'' (COFI). In this scenario, dark sector is described by a conformal field theory (CFT) at high energy scales, while at low energy confinement occurs and bound states form. One of these bound states can play the role of a dark matter candidate. In the ``fermionic" version of the COFI scenario considered in this paper, dark matter production occurs through feeble interactions with the SM sector via the so-called neutrino portal of the form $(H L) \mathcal O_\cft$. We showed that the COFI scenario provides viable stable or metastable dark matter candidates, satisfying all current phenomenological and observational constraints.
Notably, the fermionic dark matter candidate in this framework can be identified with a composite analogue of sterile neutrino. We demonstrated that the COFI production mechanism opens up new regions of parameter space that are consistent with X-ray observations, which have severely constrained existing sterile neutrino dark matter models. Future X-ray observations provide a potential observational signature of our scenario.  
Furthermore, we explored the connection between the COFI mechanism and the generation of SM neutrino masses, suggesting that the same neutrino portal responsible for dark matter production could also play a role as the origin of neutrino masses. This possibility is fascinating. However, a viable realization of this idea requires additional symmetry structure in the hadronic phase of the dark sector, as we discussed in Section~\ref{sec:Dirac}. We plan to pursue model-building along these lines in future work.


\section*{Acknowledgements}

We would like to thank Roberto Contino and Bingrong Yu for useful discussions. 
MP and TY are supported in part by the NSF grant PHY-2309456.
TY is also supported by the Samsung Science and Technology Foundation under Project Number SSTF-BA2201-06. The work of SH is supported by the National Research Foundation of Korea (NRF) Grant RS-2023-00211732, by the Samsung Science and Technology Foundation under Project Number SSTF-BA2302-05, and by the POSCO Science Fellowship of POSCO TJ Park Foundation.


\bibliographystyle{JHEP}
\bibliography{bibtex}{}
\end{document}